\documentclass[10pt,a4paper,twocolumn,english,prl,aps,showpacs,floatfix,groupedaddress,superscriptaddress]{revtex4}
\usepackage[T1]{fontenc}
\usepackage[latin9]{inputenc}
\usepackage{amsmath}
\usepackage{amssymb}
\usepackage{graphicx}
\usepackage{esint}

\makeatletter

\pdfpageheight\paperheight
\pdfpagewidth\paperwidth

\providecommand{\tabularnewline}{\\}

\@ifundefined{textcolor}{}
{%
 \definecolor{BLACK}{gray}{0}
 \definecolor{WHITE}{gray}{1}
 \definecolor{RED}{rgb}{1,0,0}
 \definecolor{GREEN}{rgb}{0,1,0}
 \definecolor{BLUE}{rgb}{0,0,1}
 \definecolor{CYAN}{cmyk}{1,0,0,0}
 \definecolor{MAGENTA}{cmyk}{0,1,0,0}
 \definecolor{YELLOW}{cmyk}{0,0,1,0}
}

\@ifundefined{definecolor}
 {\usepackage{color}}{}
\@ifundefined{definecolor}{\@ifundefined{definecolor}
 {\usepackage{color}}{}
}{}\@ifundefined{definecolor}{\@ifundefined{definecolor}{\@ifundefined{definecolor}
 {\usepackage{color}}{}
}{}}{}\def\1{1\negthickspace{\rm I}}

\makeatother

\usepackage{babel}

\makeatother

\usepackage{babel}
\begin{document}

\title{Non-local order parameters for the 1D Hubbard model}

\author{Arianna Montorsi}

\affiliation{Dipartimento di Scienza Applicata e Tecnologia, Politecnico di Torino,
corso Duca degli Abruzzi 24, I-10129 Torino, Italy}

\author{Marco Roncaglia}

\affiliation{Dipartimento di Scienza Applicata e Tecnologia, Politecnico di Torino,
corso Duca degli Abruzzi 24, I-10129 Torino, Italy}

\date{\today}
\begin{abstract}
We characterize the Mott insulator and Luther-Emery phases of the
1D Hubbard model through correlators that measure the parity of spin
and charge strings along the chain. These non-local quantities order
in the corresponding gapped phases and vanish at the critical point
$U_{c}=0$. The Mott insulator consists of bound doublon-holon pairs,
which in the Luther-Emery phase turn into electron pairs with opposite
spins, both unbinding at $U_{c}$. The behavior of the parity correlators
can be captured by an effective free spinless fermion model. 
\end{abstract}

\pacs{71.10.Hf, 71.10.Fd, 05.30.Rt}

\maketitle
The Hubbard model and its extensions have been widely used to investigate
the behavior of strongly correlated electrons in several condensed
matter systems ranging from Mott insulators (MI) to high-$T_{c}$
superconducting materials. Recently, the progress in ultracold gas
experiments that use fermionic atoms trapped into optical lattices
has opened the way to the direct simulation of the Hubbard model and
the observation of the predicted MI phase \cite{MOTT}. Since the
Mott transition is of Berezinskii Kosterlitz Thouless (BKT) type,
the MI phase does not admit a \emph{local} order parameter; instead
the transition point corresponds to the vanishing of some topological
order, possibly described by appropriate nonlocal quantities \cite{RESO,NAVO}.
A progress in this direction has been achieved in the related field
of the bosonic Hubbard models, where MI and Haldane insulator phases
have been characterized by means of non-local string parameters, inspired
by the correspondence of the bosonic system with spin-1 Hamiltonians
at low energy near integer filling \cite{DBA,GIAMARCHI}. One of these
parameters is related to the parity correlator $O_{P}(r)=\langle e^{2i\pi\sum_{j=i}^{i+r}S_{z,i}}\rangle$,
with $S_{z,i}=\frac{1}{2}(n_{i}-\nu)$ measuring the parity of the
deviation of the occupation number $n_{i}$ with respect to the filling
$\nu$ in a string starting from the site $i$, ending to the site
$i+r$. The non-vanishing value of the parity parameter $O_{P}=\lim_{r\to\infty}O_{P}(r)$
in the insulating phase has been observed with \emph{in situ} imaging
in experiments on ultracold bosonic $^{87}$Rb atoms \cite{ENDRES}.

In this Letter we address the study of nonlocal string-type correlators
to inspect the gapped phases of the fermionic Hubbard model. The expected
role of antiferromagnetic (AF) correlations has so far driven the
attention mainly to the study of Haldane type string correlators;
these were found to vanish algebraically, together with $O_{P}(r)$
in the Luttinger liquid regime \cite{KCNZ}. On the other hand, in
the large Coulomb repulsion limit the Hubbard Hamiltonian at half-filling
is known to reduce to the AF Heisenberg Hamiltonian, for which the
parity string correlator reduces trivially to the identity, the wavefunction
being frozen to the sector with only one electron per site. Since
in the MI phase the number of doubly occupied sites (doublons) and
empty sites (holons) is non-vanishing at any finite value of the interaction
(as also observed experimentally \cite{MOTT}), it is reasonable to
expect that an appropriate parity parameter could characterize the
crossover from the Heisenberg to the Luttinger liquid limit, marking
the existence of the MI phase. 

The local 4-dimensional vector space on which an electron Hamiltonian
acts is typically generated by applying to the vacuum operators forming
a $su(4)$ algebra, with three Cartan generators. Consequently, we
can introduce two independent parity correlators $O_{P}^{(\nu)}$,
defined as:
\begin{equation}
O_{P}^{(\nu)}(r)=\left\langle e^{2i\pi\sum_{j=i}^{i+r}S_{z,i}^{(\nu)}(r)}\right\rangle ,\label{eq: O_P}
\end{equation}
with index $\nu=c,s$, namely the ``charge'' and ``spin'' generalizations
of the parity correlator $O_{P}(r)$. Here $S_{z,i}^{(\nu)}$ are
the spin and pseudospin operators defined respectively as $S_{z,i}^{(s)}=\frac{1}{2}(n_{i,\uparrow}-n_{i,\downarrow})$
and $S_{z,i}^{(c)}=\frac{1}{2}(n_{i}-1)$, with $n_{i\sigma}=c_{i\sigma}^{\dagger}c_{i\sigma}$,
$\sigma=\uparrow,\downarrow$, $c_{i\sigma}^{\dagger}$ creating a
fermion at site $i$ with spin $\sigma$. By means of bosonization
and DMRG analysis, we will show that each $O_{P}^{(\nu)}$ orders
in the corresponding gapped phase: MI for $\nu=c$, with open charge
gap, and Luther Emery (LE) for $\nu=s$, with open spin gap. The $O_{P}^{(\nu)}$
vanish with the gap at the BKT transition point where the correlation
length becomes infinite. 

The Hubbard model is described by the Hamiltonian
\begin{align}
\mathcal{H}= & -\sum_{\langle ij\rangle\sigma}(c_{i\sigma}^{\dagger}c_{j\sigma}+c_{j\sigma}^{\dagger}c_{i\sigma})+U\sum_{i}n_{i\uparrow}n_{i\downarrow}\label{eq:Hubbard}
\end{align}
where the overlap integral $U$ gives the on-site contribution of
Coulomb repulsion, and energy is expressed in units of the tunneling
amplitude. 

The bosonized form of the half-filled Hubbard Hamiltonian at low-energy
is known to give rise to two continuum models describing separately
the spin and charge sectors \cite{Giamarchibook}. The latter is described
by the Hamiltonian
\begin{align}
H_{c}= & \int dx\left\{ \frac{v_{c}}{2\pi}\left[K_{c}\pi\Pi_{c}^{2}+\frac{1}{K_{c}}(\partial_{x}\Phi_{c})^{2}\right]\right.\nonumber \\
 & -\left.\frac{2U}{(2\pi\alpha)^{2}}\cos(\sqrt{8}\Phi_{c})\right\} \label{eq:sine gordon charge}
\end{align}
with
\begin{equation}
v_{c}=v_{F}\left(1+\frac{U}{\pi v_{F}}\right)^{1/2}\quad K_{c}=\left(1+\frac{U}{\pi v_{F}}\right)^{-1/2}.\label{eq:vc Kc}
\end{equation}
Here $\Phi_{c}$ is the compactified boson describing the charge excitations
with velocity $v_{c}$, and $\Pi_{c}=\partial_{x}\Theta_{c}/\pi$
is its conjugate momentum ($\alpha$ is a cutoff). At the BKT transition
point $U=0$, we have $K_{c}=1$. The bosonic field in the spin sector
$\Phi_{s}$ is governed by equations which can be obtained from (\ref{eq:sine gordon charge})
and (\ref{eq:vc Kc}) by replacing $U\to-U$ and $c\to s$. The spin-charge
transformation $c_{j\downarrow}\to(-1)^{j}c_{j\downarrow}^{\dagger}$,
that implies $S_{z,j}^{(c)}\to S_{z,j}^{(s)}$, in the present bosonization
analysis corresponds simply to the change $\Phi_{c}\leftrightarrow\Phi_{s}$.
In fact, we have used the continuum prescriptions used in Ref. \cite{Giamarchibook}
where $S^{z}(x)=\frac{\partial_{x}\Phi_{s}(x)}{\sqrt{2}\pi}$ and
$J^{z}(x)=\frac{\partial_{x}\Phi_{c}(x)}{\sqrt{2}\pi}$.

For $U>0$, we get $K_{s}>1$: the cosine term in $H_{s}$ is (marginally)
irrelevant and the spin excitations are gapless and governed by an
ordinary Gaussian model. Meanwhile, $K_{c}<1$ and a charge gap is
generated by the relevant cosine term in $H_{c}$. As a consequence,
the field $\Phi_{c}$ is pinned in one of the classical minima of
the cosine term, i.e. $\Phi_{c}=\frac{2\pi m}{\sqrt{8}}$, $m\in\mathbb{Z}$,
while $\Phi_{s}$ does not order. For $U<0$, just the same occurs
with inverted roles $\Phi_{c}\leftrightarrow\Phi_{s}$. In the continuum
limit one can realize that the parity operators become \cite{GIAMARCHI,Nakamura03}
\begin{align*}
O_{P}^{(\nu)}(r) & \approx\langle\cos[\sqrt{2}\Phi_{\nu}(r)]\cos[\sqrt{2}\Phi_{\nu}(0)]\rangle.
\end{align*}
Hence in the MI phase at $U>0$, $O_{P}^{(c)}$ turns out to be non
vanishing. In the $U<0$ case instead the LE phase is characterized
by nonzero $O_{P}^{(s)}$. The two Haldane type string correlators
$O_{S}^{(\nu)}(r)=\langle S_{z,i}^{\nu}e^{2i\pi\sum_{j=i}^{i+r}S_{z,i}^{(\nu)}(r)}S_{z,i+r}^{(\nu)}\rangle$
give instead $O_{S}^{(\nu)}(r)\approx\langle\sin[\sqrt{2}\Phi_{\nu}(r)]\sin[\sqrt{2}\Phi_{\nu}(0)]\rangle$
where the same argument suggests that these are both asymptotically
vanishing in the two gapped phases. From the above derivation, we
can conjecture that a necessary and sufficient condition for having
an asymptotically non vanishing charge (spin) parity correlator in
the Hubbard model is the opening of a gap in the charge (spin) sector,
so that $O_{P}^{(\nu)}$ do configure as order parameters for the
gapped phases of the Hubbard model. 

\begin{figure}
\includegraphics[scale=0.2]{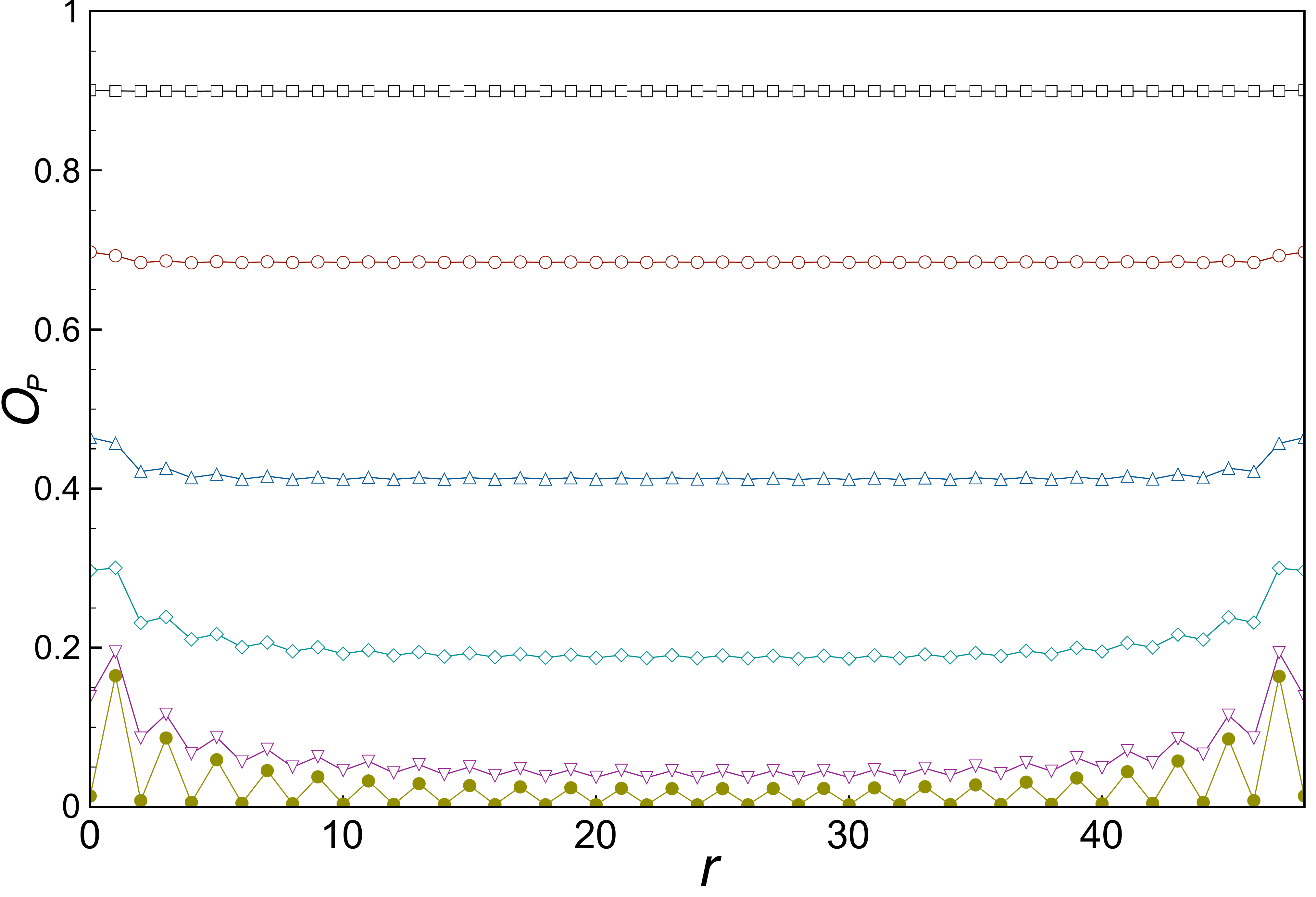}

\caption{Parity correlator $O_{P}(r)$ for a periodic chain with $L=50$ as
a function of the string length $r$. The sequences of data refer
to $U=0.1,$1.0, 2.0, 3.0, 5.0, 10.0 (in ascending order). \label{fig:OP_L50}}

\end{figure}

Below we support our previous argument providing a quantitative estimation
of the parity string parameter in the MI phase. This is achieved by
means of numerical analysis using the density matrix renormalization
group (DMRG) algorithm on finite size chains with periodic boundary
conditions (PBC's). The analysis requires very precise and reliable
data; in fact, the computing effort is significant due to both the
slowdown caused by PBC's and the high sensitivity of the correlations
contained in $O_{P}^{(\nu)}(r)$ with respect to numerical errors.
Hence we have chosen to consider chain sizes from $L=10$ to $L=50$
and $1024$ DMRG states. The curves of $O_{P}^{(c)}(r)$ plotted in
Fig.\ref{fig:OP_L50} for $L=50$ evidence clearly a fast convergence
to the asymptotic values for high interactions as well as a progressive
increase of the parity order with $U$.
 The presence of two sequences for even and odd $r$ that tend toward
the same asymptotic limit also signals that the spin parity correlator
$O_{P}^{(s)}(r)=(-1)^{r}O_{P}^{(c)}(r)$ has a uniform part $[O_{P}^{(s)}(2r+1)+O_{P}^{(s)}(2r)]/2$
that goes smoothly to zero for $U>0$. The opposite mechanism holds
for negative values of the interaction. 

Exactly at $U=0$ both parity orders are absent and $O_{P}^{(c)}(r)=O_{P}^{(s)}(r)$
as required by the spin-charge symmetry. Here, an analytic calculation
of $O_{P}^{(\nu)}(r)$ can be performed independently for both spin
species by using the Wick theorem and evaluating Toeplitz determinants.
An estimation of the asymptotic behavior gives $O_{P}^{(c)}(r)\sim r^{-1}$
at $U=0$ \cite{Abanov11}. 

\begin{figure}
\includegraphics[scale=0.34]{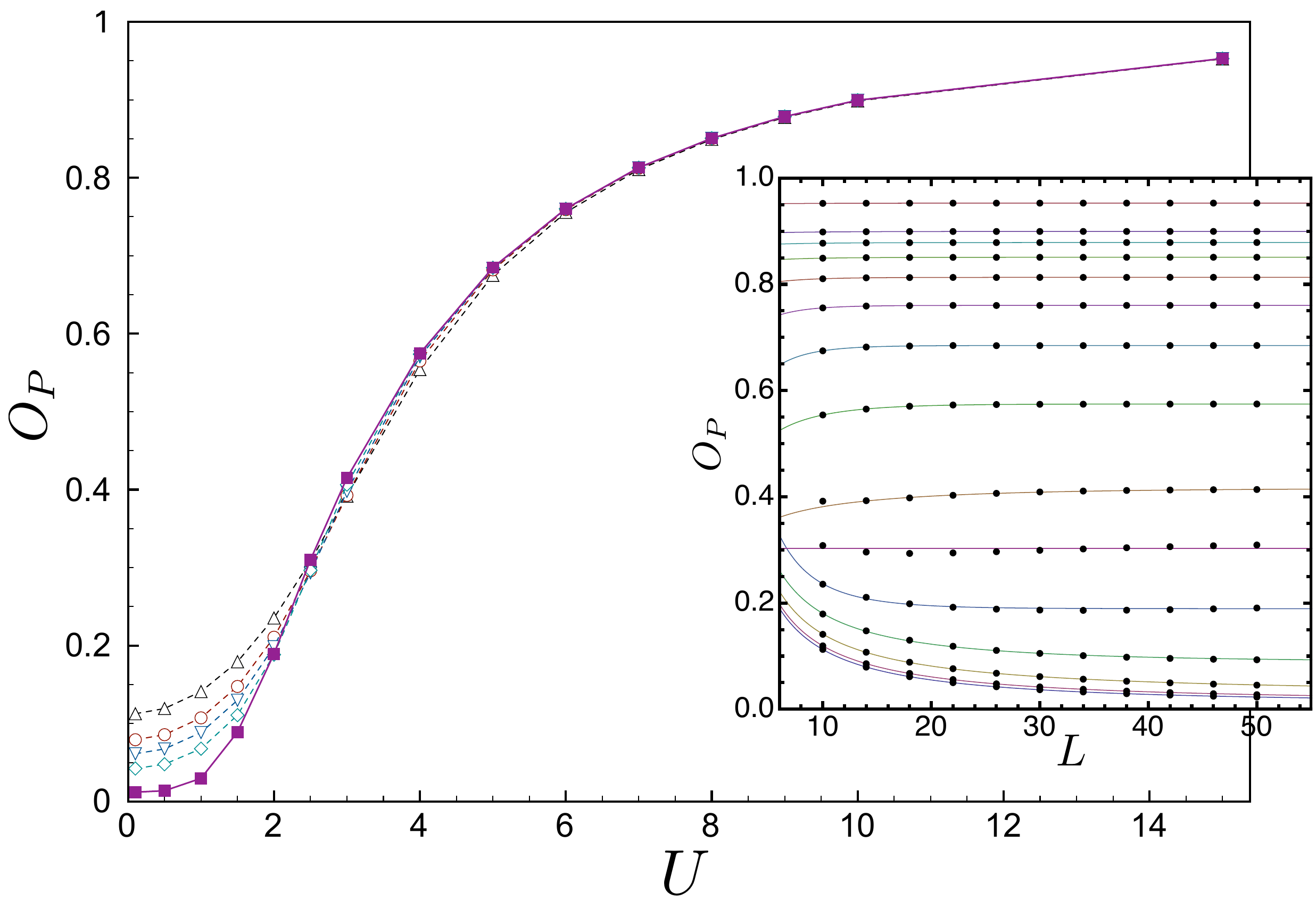}\caption{The charge parity order parameter measured at half chain $O_{P}^{(c)}(L/2)$
as a function of the local interaction $U$. We have considered PBC's
and finite chain lengths from $L=10$ to $L=50$ in step of 4, a sequence
for which the ground state of the Hubbard model is unique. We have
plotted the curves for $L=10$ (up triangles), $L=14$ (circles),
$L=18$ (down triangles), $L=26$ (rhombs). The filled squares represent
the finite size scaling values to $L=\infty$ obtained by the fits
shown in the inset. \label{fig:OP vs U} }
\end{figure}

We have explicitly evaluated the order parameter $O_{P}^{(c)}$ in
the MI phase and plotted it in Fig.\ref{fig:OP vs U} for several
values of $U$. The asymptotic values have been extrapolated from
the finite-size scaling of the quantity $O_{P}^{(c)}(L/2)$ in a periodic
chain of length $L$. For the fits, we have made use of functions
$O_{P}(r)=O_{P}+Ar^{-\gamma}e^{-r/\xi}$ obtaining a good convergence.
Interestingly, as evidenced in the inset of Fig.\ref{fig:OP vs U},
for small $U$ we get $\gamma=1$ and $A>0$, and for strong interactions
we obtain $\gamma=1/2$ and $A<0$; while for intermediate values
the best fit seems to be a combination of the two functions.

The non vanishing of $O_{P}^{(c)}$ implies the existence of bound
doublon-holon pairs; their correlation length increases by decreasing
$U$ becoming infinite at the transition, when pairs finally unbind.
The quasi long-range AF order of the MI phase suggests that such pairs
are diluted in an AF background of single electrons. The spin-charge
transformation that maps positive $U$ Hamiltonian at half-filling
into negative $U$ case at zero magnetization allows to extend the
same type of analysis to the LE phase, which is then characterized
at any filling by bound pairs of single electrons with opposite spin. 

Based on the above scenario, we construct an approximation scheme
that aims at isolating the relevant degrees of freedom (charges) to
describe the actual role of $O_{P}^{(\nu)}$ in the Hubbard model.
Since the operator $e^{i\pi n_{j}}=(-1)^{n_{j}}$ changes sign whenever
the site $j$ is singly occupied, no matter its spin orientation,
we choose to represent the original electronic creation operators
$c_{i\sigma}^{\dagger}$ in terms of a spinless fermion $f_{i}^{\dagger}$
and Pauli operators $\sigma_{i}^{a}$, $a=x,y,z$ acting on a spin
part. The mapping, schematized in Table \ref{tab:Mapping}, 
\begin{table}[b]
\begin{tabular}{|c|c|c|c|c|}
\hline 
spinful fermion & $|0\rangle$ & $|\uparrow\rangle$ & $|\downarrow\rangle$ & $|\uparrow\downarrow\rangle$\tabularnewline
\hline 
\hline 
~ spinless fermion $\otimes$ $\sigma$-spin ~  & $|0\rangle$$|+\rangle$ & $|1\rangle$$|+\rangle$ & $|1\rangle$$|-\rangle$ & $|0\rangle$$|-\rangle$\tabularnewline
\hline 
\end{tabular}\caption{Mapping from electrons to spinless fermions and $\sigma$-spins. \label{tab:Mapping}}
\end{table}
is identified by the unitary transformation
\begin{align*}
c_{i\uparrow}^{\dagger} & =c_{i\uparrow}^{\dagger}(1-n_{i\downarrow})+c_{i\uparrow}^{\dagger}n_{i\downarrow}=f_{i}^{\dagger}P_{i}^{+}+(-1)^{i}f_{i}P_{i}^{-}\\
c_{i\downarrow}^{\dagger} & =c_{i\downarrow}^{\dagger}(1-n_{i\uparrow})+c_{i\downarrow}^{\dagger}n_{i\uparrow}=(f_{i}^{\dagger}-(-1)^{i}f_{i})\sigma_{i}^{-}
\end{align*}
with $P_{i}^{\pm}=\frac{1\pm\sigma_{i}^{z}}{2}$. Interestingly, the
interaction term for the $c$-fermions simply becomes a chemical potential
shift for $f$-fermions, namely $U\sum_{i}n_{i\uparrow}n_{i\downarrow}=U(N-\sum_{i}n_{i}^{f})/2$,
where $N=\sum_{i,\sigma}n_{i\sigma}$. According to this picture,
the spin and pseudospin operators are $\mathbf{S}_{j}^{(s)}=f_{j}^{\dagger}f_{j}\boldsymbol{\sigma}_{j}$
and $\mathbf{S}_{j}^{(c)}=f_{j}f_{j}^{\dagger}\boldsymbol{\sigma}_{j}$;
conversely, we have $\boldsymbol{\sigma}_{j}=\mathbf{S}_{j}+\mathbf{J}_{j}$.

After the mapping the model in Eq.(\ref{eq:Hubbard}) becomes
\begin{align}
\mathcal{H}= & -\sum_{ij}\left[f_{i}^{\dagger}f_{j}Q_{ij}-2(-1)^{i}f_{i}^{\dagger}f_{j}^{\dagger}R_{ij}+\mathrm{H.c.}\right]\nonumber \\
 & +\frac{U}{2}\left(N-\sum_{i}f_{i}^{\dagger}f_{i}\right),\label{eq:Hubb mapped}
\end{align}
where $Q_{ij}=(\boldsymbol{\sigma}_{i}\cdot\boldsymbol{\sigma}_{j}+1)/2$
is just the swap operator in the $\sigma$-spin state and $P_{ij}^{(S)}=(1-\boldsymbol{\sigma}_{i}\cdot\boldsymbol{\sigma}_{j})/4$
is the projector onto the singlet. Notice that (\ref{eq:Hubb mapped})
is invariant under global $\sigma$-spin rotations.

The form (\ref{eq:Hubb mapped}) for the Hubbard model holds in arbitrary
dimension, and its terms are quadratic with respect to $f$-fermions.
Since $O_{P}^{(\nu)}$ can be entirely expressed in terms of $f_{i}$,
a possible strategy consists on tracing out the $\sigma$-spins by
some mean-field approximation. In fact, exploiting the symmetries
of the Hubbard model one can easily realize that $\langle Q_{ij}\rangle=1/2$
is an exact identity on the states on which the hopping term in (\ref{eq:Hubb mapped})
is non-vanishing. Moreover, we set the parameter $\alpha\equiv\langle R_{ij}\rangle$
in a phenomenological way by equating the ground state (GS) energy
obtained from the spinless quadratic model with the exact energy coming
from the Bethe-Ansatz solution \cite{LIWU}. Within this approximation
Eq.(\ref{eq:Hubb mapped}) is diagonalized in Fourier space, obtaining

\[
H=\sum_{k\in BZ}\Lambda_{k}\left[\eta_{k}^{\dagger}\eta_{k}-\frac{1}{2}\right]+\frac{U(2N-L)}{4}\:,
\]
with spectrum $\Lambda_{k}=-\cos k+\sqrt{16\alpha^{2}\cos^{2}k+U^{2}/4}$
and $\eta_{k}$ are the new fermionic modes. In the thermodynamical
limit (TL), the energy density $e_{GS}$ at half-filling $\nu=1$
is given by $e_{GS}=\frac{U}{4}-\frac{1}{2\pi}\int_{-\pi/2}^{\pi/2}\mathrm{d}k\:\sqrt{16\alpha^{2}\cos^{2}k+U^{2}/4}.$
It is interesting to observe that the model is gapless only for $U=0$,
where for $\alpha=1$ $e_{GS}$ assumes the exact value of the non-interacting
case. For $U>0$ the number of singly occupied states $\nu_{f}$ is
increasing and the pair-singlet states start to interact. 

We are interested in calculating the parity operator $O_{P}^{(c)}(r)=\langle e^{i\pi\sum_{j=i}^{i+r}(n_{j}^{f}-1)}\rangle$,
that can be rewritten as 
\begin{align*}
O_{P}^{(c)}(r) & =\left\langle \prod_{j=i}^{i+r}(2f_{j}^{\dagger}f_{j}-1)\right\rangle =\left\langle \prod_{j=i}^{i+r}A_{j}B_{j}\right\rangle 
\end{align*}
having defined $A_{j}=(f_{j}^{\dagger}+f_{j})$ and $B_{j}=(f_{j}-f_{j}^{\dagger})$.
Making use of the Wick theorem, $O_{P}^{(c)}(r)$ can be expressed
as a determinant \cite{LSM61}
\begin{equation}
O_{P}^{(c)}(r)=\left|\begin{array}{ccccc}
G_{0} & G_{1} & G_{2} & \cdots & G_{i,i+r}\\
-G_{1} & G_{0} & -G_{1} & \cdots\\
G_{2} & G_{1} & G_{0} & \cdots\\
\vdots & \vdots & \vdots & \ddots & \cdots\\
G_{i+r,i} &  &  & \vdots & G_{0}
\end{array}\right|=\det(\mathbf{G}).\label{eq:Op_determinant}
\end{equation}
where $\mathbf{G}$ is a \emph{block Toeplitz matrix} of dimension
$(r+1)\times(r+1)$, whose entries are the one-body correlation functions
$G_{r}=\langle(f_{j}^{\dagger}-f_{j})(f_{j+r}^{\dagger}+f_{j+r})\rangle$,
whose expressions in the TL are
\[
G_{r}=\begin{cases}
\frac{U}{2\pi}\int_{0}^{\pi}\mathrm{d}k\:\frac{\cos(kr)}{\sqrt{16\alpha^{2}\cos^{2}(k)+U^{2}/4}}, & r\;\mathrm{even}\\
(-1)^{j}\frac{4\alpha}{\pi}\int_{0}^{\pi}\mathrm{d}k\:\frac{\cos(kr)\cos(k)}{\sqrt{16\alpha^{2}\cos^{2}(k)+U^{2}/4}}, & r\;\mathrm{odd}
\end{cases}
\]
with the property that $G_{0}=2\nu_{f}-1$, $G_{r}=2\mathrm{Re}(\langle f_{j}^{\dagger}f_{j+r}\rangle)$
for $r$ even and $G_{r}=2\mathrm{Re}(\langle f_{j}^{\dagger}f_{j+r}^{\dagger}\rangle)$
for $r$ odd. The blocks in (\ref{eq:Op_determinant}) are of size
$2\times2$. We must distinguish the cases of $r$ even or odd, since
they give rise to two different sequences. In particular, here we
stick to the case $r$ odd, where the block matrix is of even dimension. 

\begin{figure}
\includegraphics[scale=0.23]{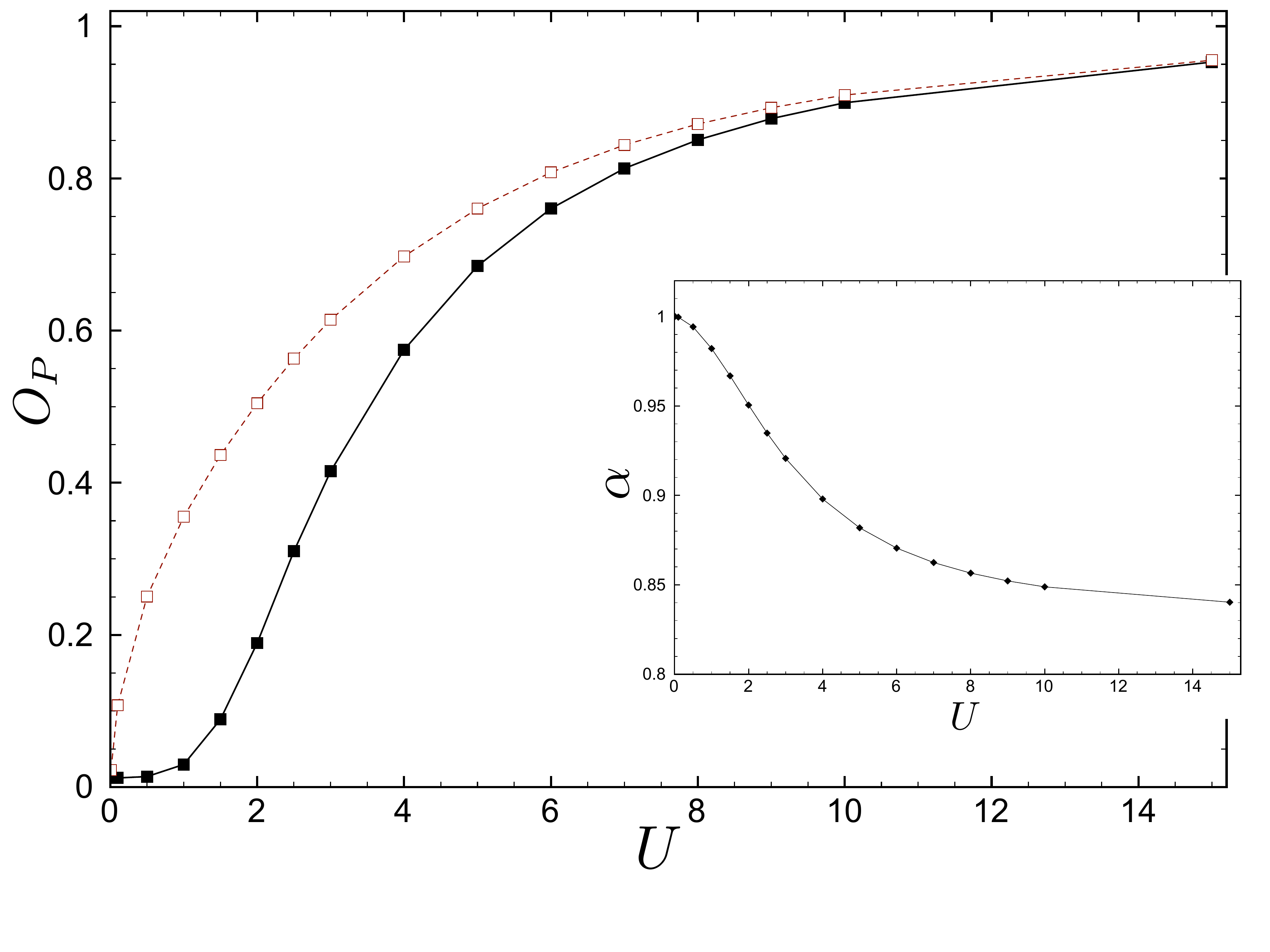}

\caption{The parity order $O_{P}^{(c)}$ calculated according to Eq.(\ref{eq:Op_determinant})
as a function of $U$ (dashed line) in the MI phase of the effective
model (\ref{eq:Hubb mapped}) with $\alpha$ as in the inset figure,
determined by tuning the spinless fermion energy to the exact Hubbard
value. The results are compared with the numerical curve (continuous
line) obtained for the Hubbard model and shown in Fig.\ref{fig:OP vs U}.\label{fig:Op Hf}}

\end{figure}

The analytical calculation of $O_{P}^{(c)}$ in the TL for some values
of $U$ in the $f$-fermion approximation yields to the curve plotted
in Fig.\ref{fig:Op Hf}, evidentiating the expected non vanishing
of the charge parity order for $U>0$. The parameter $\alpha$ has
been determined by requiring $e_{GS}(U,\alpha)=e_{ex}(U)$, where
$e_{ex}$ is the exact result \cite{LIWU}.
Remarkably, such equality admits a solution for every $U$, which
belongs to a narrow interval below $\alpha=1$, as shown in the inset
of Fig.\ref{fig:Op Hf}. This means that in the pair-creation processes
in (\ref{eq:Hubb mapped}) the $\sigma$-spin state is very close
to the singlet. In the limit $U\gg1$ the energy becomes $e_{GS}(\alpha)\approx-4\alpha^{2}/U$
that gives $\alpha(U\to\infty)=\sqrt{\log2}\approx0.83$, by comparison
with the energy density of the Heisenberg model coming from the large-$U$
expansion of the Hubbard model at $\nu=1$. The result for $O_{P}^{(c)}$
is also quantitatively in accordance with the DMRG data in the large
$U$ region, where our assumptions on the AF nature of short-ranged
correlations \cite{noi10} are more justified. 

In conclusion, our work unveils that two (charge and spin) hidden
parity string correlators play the role of order parameters for the
gapped phases of the Hubbard model. In the bosonization approach these
are found to be asymptotically finite only in the corresponding gapped
MI and LE phases, and vanish with the gap at BKT transition point.
The result is cleanly confirmed by DMRG numerical analysis. The emerging
physical scenario is that of an insulator in which bound pairs of
doublons and holons move in a AF background of single electrons. In
the LE regime the role of doublons and holons and that of up and down
electrons are exchanged. The picture allows to derive an effective
free spinless fermion model which captures correctly the presence
of non local order, and its vanishing at the transition. The spinless
model is exact in the limit of large $U$ thus complementing the standard
strong-coupling description with $t-J$ model.

The parity order is suitable for experimental detection by high resolution
imaging \cite{ENDRES} in ultracold Fermi gases. Possibly, some of
the features described here could persist in two dimensions, where
the localization of bound pairs could take place along one dimensional
stripes. The scenario is quite suggestive also from the perspective
of high-$T_{c}$ materials: the presence of bound doublon-hole pairs
in the undoped insulator could play a role upon doping in the onset
of the superconducting phase. 

The present analysis could be further exploited to extended Hubbard
models \cite{DOMO,AAA}, to describe other topologically ordered phases;
noticeably, the fully gapped phase characterized by non vanishing
charge and spin gaps should correspond to the non vanishing of both
$O_{P}^{(\nu)}$'s. Work is in progress along these lines.

M.R. acknowledges support from the EU-ERC project no. 267915 (OPTINF)
.

\end{document}